\documentclass[twocolumn,showpacs,preprintnumbers,amsmath,amssymb,superscriptaddress]{revtex4}

\usepackage{graphicx}
\usepackage{dcolumn}
\usepackage{bm}

\begin{document}
\def\Journal#1#2#3#4{{#1}{\bf #2}, #3 (#4)}
\def\bi#1{#1}
\def\NCA{ Nuovo Cimento}
\def\NIM{ Nucl. Instrum. Methods}
\def\NIMA{{ Nucl. Instrum. Methods in Phys. Research Sect.} A}
\def\NIMB{{ Nucl. Instrum. Methods} B}
\def\NPB{{ Nucl. Phys.} {\bf B}}
\def\NPBP{{ Nucl. Phys.} B (Proc. Suppl.) }
\def\NPA{{ Nucl. Phys.} {\bf A}}
\def\PLB{{ Phys. Lett.}  {\bf B}}
\def\PRL{ Phys. Rev. Lett. }
\def\PRD{{ Phys. Rev.} {\bf D}}
\def\PRC{{ Phys. Rev.} {\bf C}}
\def\PRA{{ Phys. Rev.} A}
\def\PR{{ Phys. Rev. }}
\def\PRP{{ Phys. Rep. }}
\def\ZPC{{ Z. Phys.} C}
\def\ZPA{{ Z. Phys.} A}
\def\ZP{{ Z. Phys. }}
\def\MPA{{ Mod. Phys. Lett. A}}
\def\PS{{ Physics Scripta }}
\def\PAN{ Phys. Atom. Nucl.}
\def\ARAA{{ Ann. Rev. Astr. Astroph.}}
\def\ARNP{{ Ann. Rev. Nucl. Part. Phys.}}
\def\GCA{ Geochim. Cosmochim Acta}
\def\PNPP{ Prog. Nucl. Part. Phys.}
\def\PTP{ Prog. Theo. Phys.}
\def\PTPS{ Prog. Theo. Phys. Suppl.}
\def\APJ{ Ap.J. }
\def\AA{ Astron. Astroph.}
\def\AAS{ Astron. Astroph. Suppl. Ser.}
\def\RPP{ Rep. Prog. Phys.}
\def\RMP{ Rev. Mod. Phys.}
\def\APP{ Astroparticle Physics }
\def\EPL{ Europhys. Lett.}
\def\IJMP{ Int. Journal of Modern Physics}
\def\JPG{ Journal of Physics G}
\def\NAT{ Nature}
\def\SCI{ Science }
\def\JETP{ JETP Lett. }
\def\JHEP{ J. High Energy Physics}
\def\ANDT{ Atom. Nucl. Dat. Tab.}
\def\BASI{ Bull. Astr. Soc. India}
\def\BAPS{ Bull. Am. Phys. Soc.}
\def\SJNP{ Sov. Journal of Nucl. Phys.}
\def\ZETZ{ Zh. Eksp Teor. Fiz.}

\def\nubar{$\bar{\nu}$}
\def\nuebar{$\bar{\nu}_e$}
\def\nue{$\nu_e$ }
\def\numubar{$\bar{\nu}_{\mu}$}
\def\nutaubar{$\bar{\nu}_{\tau}$}
\def\ccbar{${\overline{\text{CC}}}$}

\title{Electron Antineutrino Search at the Sudbury Neutrino Observatory}

%
\newcommand{\ubc}{Department of Physics and Astronomy, University of
British Columbia, Vancouver, BC V6T 1Z1 Canada}
\newcommand{\bnl}{Chemistry Department, Brookhaven National
Laboratory,  Upton, NY 11973-5000}
\newcommand{\carleton}{Ottawa-Carleton Institute for Physics, 
Department of Physics, Carleton University, Ottawa, Ontario K1S 5B6 
Canada}
\newcommand{\uog}{Physics Department, University of Guelph,
Guelph, Ontario N1G 2W1 Canada}
\newcommand{\lu}{Department of Physics and Astronomy, Laurentian
University, Sudbury, Ontario P3E 2C6 Canada}
\newcommand{\lbnl}{Institute for Nuclear and Particle Astrophysics and
Nuclear Science Division, Lawrence Berkeley National Laboratory, 
Berkeley, CA 94720}
\newcommand{\lbla}{ Lawrence Berkeley National Laboratory, Berkeley, CA}
\newcommand{\lanl}{Los Alamos National Laboratory, Los Alamos, NM 87545}
\newcommand{\lanla}{Los Alamos National Laboratory, Los Alamos, NM}
\newcommand{\oxford}{Department of Physics, University of Oxford,
Denys Wilkinson Building, Keble Road, Oxford, OX1 3RH, UK}
\newcommand{\penn}{Department of Physics and Astronomy, University of
Pennsylvania, Philadelphia, PA 19104-6396}
\newcommand{\queens}{Department of Physics, Queen's University,
Kingston, Ontario K7L 3N6 Canada}
\newcommand{\uw}{Center for Experimental Nuclear Physics and Astrophysics,
and Department of Physics, University of Washington, Seattle, WA 98195}
\newcommand{\triumf}{TRIUMF, 4004 Wesbrook Mall, Vancouver, BC V6T 2A3, Canada}
\newcommand{\ralsuss}{Rutherford Appleton Laboratory, Chilton, Didcot,
Oxon, OX11 0QX, and University of Sussex, Physics and Astronomy Department,
Brighton BN1 9QH, UK}
\newcommand{\uta}{Department of Physics, University of Texas at Austin,
Austin, TX 78712-0264}
\newcommand{\iusb}{Department of Physics and Astronomy, Indiana 
University, South Bend, IN}
\newcommand{\fnal}{Fermilab, Batavia, IL}
\newcommand{\uo}{Department of Physics and Astronomy, University of 
Oregon, Eugene, OR}
\newcommand{\rcnp}{Research Center for Nuclear Physics and Osaka 
University, Osaka, Japan}
\newcommand{\slac}{Stanford Linear Accelerator Center, Menlo Park, CA}
\newcommand{\mac}{Department of Physics, McMaster University, Hamilton, ON}
\newcommand{\doe}{US Department of Energy, Germantown, MD}
\newcommand{\lund}{Lund University, Sweden}
\newcommand{\mpi}{Max-Planck-Institut for Nuclear Physics, Heidelberg, Germany}
\newcommand{\uom}{Ren\'{e} J.A. L\'{e}vesque Laboratory, 
Universit\'{e} de Montr\'{e}al, Montreal, PQ}
\newcommand{\cwru}{Department of Physics, Case Western Reserve 
University, Cleveland, OH}
\newcommand{\pnnl}{Pacific Northwest National Laboratory, Richland, WA}
\newcommand{\uc}{Department of Physics, University of Chicago, Chicago, IL}
\newcommand{\mitt}{Department of Physics, Massachusetts Institute of 
Technology, Cambridge, MA }
\newcommand{\ucsd}{Department of Physics, University of California at 
San Diego, La Jolla, CA }
\newcommand{\lsu}{Department of Physics and Astronomy, Louisiana 
State University, Baton Rouge, LA }

\affiliation{	\ubc	}
\affiliation{	\bnl	}
\affiliation{	\carleton	}
\affiliation{	\uog	}
\affiliation{	\lu	}
\affiliation{	\lbnl	}
\affiliation{	\lanl	}
\affiliation{	\oxford	}
\affiliation{	\penn	}
\affiliation{	\queens	}
\affiliation{	\ralsuss	}
\affiliation{	\triumf	}
\affiliation{	\uw	}

\author{	B.~Aharmim	}			\affiliation{ 
	\lu	}
\author{	S.N.~Ahmed	}			\affiliation{ 
	\queens	}
\author{	E.W.~Beier	}			\affiliation{ 
	\penn	}
\author{	A.~Bellerive	}			\affiliation{ 
	\carleton	}
\author{	S.D.~Biller	}			\affiliation{ 
	\oxford	}
\author{	J.~Boger	} 
	\altaffiliation{Present Address: \doe}	\affiliation{	\bnl 
	}
\author{	M.G.~Boulay	}			\affiliation{ 
	\lanl	}
\author{	T.J.~Bowles	}			\affiliation{ 
	\lanl	}
\author{	S.J.~Brice	} 
	\altaffiliation{Present address: \fnal}	\affiliation{	\lanl 
	}
\author{	T.V.~Bullard	}			\affiliation{ 
	\uw	}
% ...

\collaboration{SNO Collaboration}
\noaffiliation

\date{\today}

\begin{abstract}

Upper limits on the \nuebar\, flux at the
Sudbury Neutrino Observatory 
have been set based on the \nuebar\, charged-current
reaction on deuterium.
The reaction produces a positron and two neutrons
in coincidence. This distinctive signature allows a
search with very low background for \nuebar's\, from the Sun
and other potential sources.
Both differential and integral limits on the \nuebar\, flux have been placed
in the energy range from 4 -- 14.8~MeV.
For an energy-independent $\nu_{e}$ $\rightarrow$ $\bar{\nu}_{e}$ conversion
mechanism, the integral limit on the flux of solar \nuebar's\, in the energy range from 4 -- 14.8~MeV is found to be
$\Phi_{\bar{\nu}_{e}} \le 3.4 \times 10^4\mathrm{cm}^{-2} \mathrm{s}^{-1} $ (90\% C.L.),
which corresponds to 0.81\%  of the standard solar model $^8$B $\nu_e$ flux
of $5.05\times 10^6\mathrm{cm}^{-2} \mathrm{s}^{-1} $, and is 
consistent with the more sensitive limit from KamLAND in the 8.3 
-- 14.8~MeV range of $3.7\times 10^2\mathrm{cm}^{-2} \mathrm{s}^{-1} $\
 (90\% C.L.).
In the energy range from 4 -- 8~MeV, a search for \nuebar's\, 
is conducted using coincidences in which only the two neutrons are detected.
Assuming a \nuebar\, spectrum for the neutron induced fission of naturally 
occurring elements, a flux limit of
$\Phi_{\bar{\nu}_{e}} \le 2.0 \times 10^6\mathrm{cm}^{-2} \mathrm{s}^{-1} $ 
(90\% C.L.) is obtained. 

\end{abstract}


\maketitle

\section{Introduction}
This paper presents results from a search for 
\nuebar's \, with the Sudbury Neutrino Observatory (SNO) 
via the charged-current reaction (\ccbar)
on deuterons :
\begin{displaymath}
\bar{\nu}_e + d \rightarrow e^+ + n + n - 4.03~\mathrm{MeV}.
\end{displaymath}
The distinctive signature of a positron in coincidence with 
two neutrons allows SNO to search for \nuebar's \, 
with very low background.
By means of ($n,n$)-coincidence detections, SNO has 
sensitivity to \nuebar's \, with energies above the reaction threshold 
of 4.03~MeV. For 
coincidences involving a $e^+$, SNO is sensitive to \nuebar's\, above a
threshold of 4~MeV + E$_{\mathrm{recoil}}^{\mathrm{thr}}$, where E$_{\mathrm{recoil}}^{\mathrm{thr}}$ is the 
analysis threshold applied to the recoil positron.
We present results for integral \nuebar\, flux limits 
in the neutrino energy range from 4 -- 14.8~MeV under the assumption that 
\nuebar's originate from a $^8$B spectrum, and below 8~MeV
under the assumption that \nuebar's originate from a fission spectrum.
Differential limits on the \nuebar\, flux have been placed,
independent of any particular spectral assumptions.\\
\indent As a solution to the solar neutrino problem \cite{sol76},
analyses of global solar neutrino data favor matter-enhanced
neutrino oscillations with mixing
parameters in the Large Mixing Angle (LMA) region \cite{snodn,glo02}.
Studies of solar neutrino data
have demonstrated that approximately two-thirds of $^8$B solar neutrinos 
convert to active flavors other than \nue before reaching    
Earth~\cite{SKsolar,snocc,snonc,snodn,egu03,sno03}. 
The deficit of reactor antineutrinos reported by 
the Kamioka Liquid scintillator Anti-Neutrino Detector
(KamLAND) experiment \cite{egu03} 
supports the LMA solution to the solar neutrino problem under the 
explicit assumption of CPT conservation in the neutrino sector.
   More generally the spin flavor precession (SFP) mechanism
   \cite{sfp02,akh03} or neutrino decay \cite{nudec} could contribute 
   to the observed neutrino flavor
   transformation at a sub-dominant level by converting a small fraction
   of solar $\nu_e$ to $\bar{\nu}_e$.  In SFP models, neutrinos are
   assumed to have a transition magnetic moment of the order of 
  10$^{-11}$ Bohr magnetons \cite{akh03} if they are of Majorana type.
   Solar magnetic
   fields, which are known to be time dependent, 
couple to this magnetic moment to convert $\nu_e$ into
   $\bar{\nu}_e$ with a combination of neutrino flavor oscillations
   and SFP mechanisms. Neutrino
   decay models allow heavier neutrino mass eigenstates to decay into
   light \nubar \,\cite{nudec}. \\
\indent 

Nuclear fission \nuebar\, spectra peak
at low energies and fall approximately exponentially with energy,
and have negligible intensity above 8~MeV \cite{vog81}. 
Positrons produced in \ccbar\, reactions from fission \nuebar's are too low in 
energy to be detected by the present \nuebar\, analysis. 
However, by conducting an analysis involving
only ($n,n$)-coincidences it is possible to study the energy region from 4
-- 8 MeV, providing some sensitivity to \nuebar's that might originate from
naturally occurring neutron induced fission sources.
Because the fission spectrum
is significant in this energy region, whereas the $^8$B spectrum is peaked
at higher energies, a separate analysis of this region for \nuebar's
originating from fission is performed. The expected flux from man-made reactors
would provide a negligible contribution to this analysis. 

\indent Additional sources of \nuebar's are cosmic ray interactions in the upper 
atmosphere and the diffuse supernovae background. The flux of these types
of \nuebar's are dominated by \nuebar\, energies above the 4 -- 14.8~MeV 
energy range investigated in the present analysis and hence are only
addressed peripherally by the differential limits below 15~MeV.

\section{Experimental Data}

SNO is an imaging water Cherenkov detector located at a depth of 
6010 m of water equivalent in the Inco, Ltd.\ Creighton mine near Sudbury, 
Ontario, Canada.
SNO detects neutrinos using an ultra-pure heavy water target contained in a 
transparent acrylic spherical shell 12~m in
diameter.  
Cherenkov photons generated in the heavy 
water are detected by 9456 photomultiplier tubes (PMTs) mounted on a stainless 
steel geodesic sphere 17.8 m in diameter.  
The geodesic sphere is immersed in 
ultra-pure light water to provide shielding from radioactivity in both the PMT 
array and the cavity rock.
The SNO detector has been described in detail in~\cite{sno}. 

The data reported here were recorded between November 2, 1999 and May 28, 2001 and
span the entire first phase of the experiment, in which only D$_2$O was 
present in the sensitive volume.  In
comparison to earlier SNO analyses \cite{snocc,snonc}, a cut  
employed to remove events following muon candidates was modified
to maximize rejection of false \nuebar\, candidates. 
Any event with an assigned kinetic energy above 18 MeV and all events 
following that event within 0.5~s are rejected.
These modifications resulted in a total livetime of 305.9 live days, a reduction
of less than 0.2\% from SNO's neutral-current (NC) analysis \cite{snonc}. 

\section{Analysis}

Interactions of \nuebar's \,  with deuterons produce
a positron and two neutrons. The positron generates
a prompt Cherenkov signal, while the neutrons must first thermalize before
generating 6.25~MeV gamma rays from their capture on deuterons. The mean
neutron capture time is 42~ms and the diffusion length is $\sim 110$~cm in pure D$_2$O.
Inside the D$_2$O volume \nuebar \, events can be identified by 
a coincidence of 2 or 3 particles.

The analysis procedure consists of two steps. The first is similar 
to the SNO solar neutrino analysis and is described in \cite{snonc}. 
In this step, PMT times and hit patterns are used
to reconstruct event vertices and assign to each recoil positron or electron
a most probable 
kinetic energy, $T_{\mathrm{eff}}$.
The recoil threshold in this analysis was $T_{\mathrm{eff}} \ge$ 5~MeV, 
providing sensitivity
to positrons and neutrons from the 
 $\overline{\text{CC}}$ reaction.
No \nuebar's\, originating from the conversion of 
solar $\nu_e$'s are expected to have energies
in excess of 14.8~MeV. Hence, the 18~MeV upper limit on
recoil positron or electron energy does not remove potential solar or fission \nuebar \, candidates.
A fiducial volume was defined by requiring reconstructed event vertices to be
within $550$~cm of the center of the acrylic vessel.
This reduces both the number of externally produced background events, and 
the systematic 
uncertainties associated with optics and event reconstruction near 
the acrylic vessel.

The second step of the analysis identifies coincidences among the accepted 
events.
The size of the coincidence window, chosen to be 150~ms, 
was optimized to maximize the sensitivity to
2-fold coincidences in the presence of the background 
of accidental coincidences. Antineutrino candidates that are part
of a burst of 4 or more events of any energy are discarded. 
Even under the assumption that 100\% of the solar \nue are converted to \nuebar's, four- or higher-fold    
coincidences are $10^3$ times more likely to originate from a    
background such as atmospheric $\nu$'s or spontaneous fission of    
$^{238}$U than from a 3-fold $\bar{\nu}_e$ event in coincidence with
a random event, as determined by Monte Carlo simulations.

\subsection{Detection Efficiencies}
\begin{figure}[htb]
\includegraphics[width=3.4in]{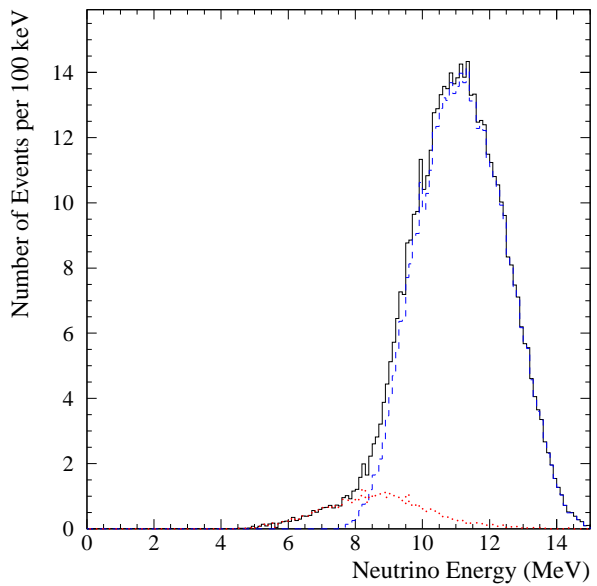}
\caption{Number of expected coincidence events as a function of the
antineutrino energy, for \nuebar's originating from a
$^8$B spectrum with a total flux of 5.05$\times 10^{6}$cm$^{-2}$s$^{-1}$.
Coincidences containing a positron and ($n,n$)-coincidences
are shown as dashed and dotted histograms, respectively. 
The solid line represents all types of detected coincidences 
as a function of \nuebar\, energy.}
\label{fig:deteff}
\end{figure}

Coincidence detection efficiencies
were determined from a simulated \nuebar \, data set. 
The Monte Carlo (MC) simulations sampled \nuebar's \, from a $^8$B energy 
spectrum \cite{ort02} with a total flux 100 times the standard 
solar model (SSM-BP00) $^8$B flux \cite{bp00}. 
The simulated data set matches the experimentally recorded data set
in duration, and correctly describes the 
detector status as a function of time.
Based on the number of simulated and extracted coincidences
inside the signal region, the
2- and 3-fold coincidence event detection efficiencies 
were found to be:
\nobreak{
\begin{eqnarray*}
\epsilon_{(e^+,n,n)} = \,\,\, 1.11 \, \pm 0.02 (\mathrm{stat.}) \, ^{+0.05} _{-0.12} (\mathrm{syst.})\, \% \\
\epsilon_{(e^+,n)} =  10.27 \, \pm  0.05 (\mathrm{stat.}) \, ^{+0.37} _{-0.94} (\mathrm{syst.})\,\% \\
\epsilon_{(n,n)} = \,\,\, 1.20 \, \pm  0.02 (\mathrm{stat.}) \, ^{+0.05} _{-0.10} (\mathrm{syst.})\, \%
\end{eqnarray*}}Among 
coincidences originating from \nuebar \,
interactions, ($e^+,n$)-coincidences are 10 times more likely to be 
detected than ($n,n$)-coincidences or ($e^+,n,n$)-coincidences.
While the ($n,n$)-coincidences can originate from \nuebar's \, with energies
as low as 4~MeV, coincidences containing a positron must originate from
\nuebar's \, with energies above 9~MeV.
Figure \ref{fig:deteff} displays the number of expected \nuebar\, events
as function of the
antineutrino energy under the assumption that \nuebar's \, originate from
a $^8$B spectrum with a total flux of 5.05$\times 10^{6}$cm$^{-2}$s$^{-1}$. 
The above coincidence detection efficiencies are consistent with the 
average $e^+$ detection efficiency estimated from Monte Carlo
$\overline{\epsilon}_{e^+} = 40.09 \,\,^{+ 1.50} _{- 4.62} \,\, (\mathrm{syst}) \pm 0.10 \,\, (\mathrm{stat})$\%, 
and an average neutron detection efficiency of 
$\overline{\epsilon}_{n}= 14.38 \pm 0.53 $\% \cite{snonc}. The neutron response and
systematic uncertainty on the response were calibrated with a $^{252}$Cf source.

\subsection{Backgrounds}
Backgrounds to the \nuebar \, search can be divided into 
two categories: coincidences caused by 
\nuebar's\, from known sources, and 
coincidences from processes other than \nuebar\, charged-current 
reactions. These are presented in Table \ref{tab_antinu_bkgd}. 
The first category has contributions from atmospheric, reactor, and
diffuse supernovae \nuebar's.
The background contribution from atmospheric \nuebar's\, is 
estimated to be 0.07 $\pm$ 0.01 coincidences. It
is derived from the ${\overline{\text{CC}}}$ 
cross sections \cite{nak02}, 
coincidence detection efficiencies, and 
a parameterized \nuebar \, spectrum extrapolated
into the energy range below 50~MeV \cite{gai89}. 
Highly energetic atmospheric \nuebar's\,
only contribute to the background if detected as ($n,n$)-coincidences, 
due to the applied upper bound on the recoil lepton energy. 
The background estimate from nuclear power reactors yields
0.019 $\pm$ 0.002 coincidences. The calculation is
based on the monthly reported actual power output of all commercial 
reactors \footnote{We would like to thank the US Nuclear Regulatory 
Commission and AECL Chalk River Laboratories for providing this information.}
within 500~km of SNO and an average reactor \nuebar\, 
spectrum \cite{vog81,sch89}. Furthermore, the \nuebar\, flux is
assumed to be reduced as a result of neutrino oscillations governed
by the best fit oscillation parameters tan$^2$($\theta$)=0.41 and 
$\Delta$m$^2$=7.1$\times 10^{-5}$~eV$^2$ \cite{sno03}.
The number of ${\overline{\text{CC}}}$ interactions from diffuse
supernovae neutrinos was estimated by combining a theoretical flux calculation \cite{and03}
with SNO's detection efficiency and contributes 
$\le$ 0.005 coincidences. The calculation is
consistent with experimental limits \cite{mal02}.
Electron antineutrinos from the decay chain of terrestrial radioisotopes 
such as $^{238}$U and $^{232}$Th, geo-antineutrinos, do not contribute to 
the background because their energies are below the threshold of the 
\ccbar\, reaction.\\
\indent  The main contributions to the second category of backgrounds originate
from atmospheric $\nu$, possible $^{238}$U fission events in the detector 
media,  and accidental coincidences.
The atmospheric $\nu$ background is estimated to account for
1.46$^{+0.49}_{-0.45}$ coincidences. Atmospheric neutrinos were sampled from a 
realistic spectrum \cite{gai89} and propagated with the neutrino-interaction 
generator NUANCE \cite{cas02}. Detailed event information, including
energy and multiplicities at the depth of SNO, was then 
processed by a full detector simulation.\\
\indent Spontaneous fission background from $^{238}$U in detector media 
was estimated to contribute less than 0.79 coincidences. 
This limit was derived from 
an inductively coupled plasma mass spectrometer (ICP-MS)
 measurement of the $^{238}$U concentration of 66$\pm17$~(fg U)/(g D$_2$O)
\footnote{The number quoted in \cite{snonc}
is smaller by a factor of four but cannot be used here 
because it represents (g U $equivalent$)/(g D$_2$O), 
which assumes equilibrium in the $^{238}$U decay chain. 
We thank L. Yang and R. Sturgeon of the
National Research Council of Canada for this measurement.}, 
for which approximately 14 spontaneous fission decays of $^{238}$U 
are expected per kton-year.
The detector response was calculated on the basis of detection efficiencies 
and a discrete probability distribution for neutron multiplicities
in spontaneous fission \cite{tur57}. 
The above estimate is an upper limit since the $^{238}$U concentration
measurement was obtained on September 4, 2003, after the data set
reported here, and after the addition of NaCl to the heavy water.  Because the
reverse osmosis purification system, which was operated while the data
accumulated for the present analysis, could not be run with NaCl in the heavy
water, the measurement is likely to include more $^{238}$U contamination 
than was present when the data were taken.\\
\indent Accidental coincidences are formed by individual events that pass 
the analysis cuts and have a random time correlation. 
Their number has been calculated as a function of the time-dependent 
singles rate in the detector and amounts to 0.13 $^{+0.06}_{-0.04}$
coincidences. This number was confirmed independently by a
direct measurement of signals following within 150~ms of a large
number of random triggers.\\
\indent The background from neutron capture reactions on oxygen,
which can produce multiple gamma-rays above the deuteron photodisintegration
threshold,
was estimated to be less than 0.05 coincidences.
It was calculated from abundances of $^{17}$O and $^{18}$O,
relative intensities of gamma rays produced in
$^{17}$O(n,$\gamma$)$^{18}$O and $^{18}$O(n,$\gamma$)$^{19}$O reactions,
the ratio ${P_{^{17}\mathrm{O},{^{18}\mathrm{O}}}}/P_\mathrm{D}$ of the
neutron capture probability on $^{17}$O and $^{18}$O and
deuterium, as well as the total number of observed neutrons. 
Instrumental backgrounds are events produced by electrical pickup or
emission of non-Cherenkov light from detector components.
Their background contribution is determined to be
$<$ 0.027 coincidences (95\% C.L.).
The number of coincidences of this type
is assessed by means of a bifurcated analysis,
which employs sets of orthogonal cuts aimed at instrumental background
rejection. 
The background from $\alpha$-capture reactions on carbon, which can produce
a neutron in coincidence with an $e^+-e^-$ pair, was found 
to be less than 1.7$\times 10^{-3}$ (90\% C.L.) coincidences.
It was estimated on the basis of
the total number of neutrons in the signal region and a MC calculation.
Other backgrounds originate from radioisotope contamination 
and can produce coincidences through
$\beta\mbox{-}\gamma$ or $\beta\mbox{-}n$ decays, 
but are found to be entirely negligible.
They are estimated on the basis of their respective radioisotope
contamination levels.\\
\indent Upper limits and uncertainties on individual backgrounds
have been combined under the assumption that they are independent.
The total background amounts to 1.68 $^{+0.93}_{-0.45}$ coincidences.

\squeezetable
\begin{table}
\begin{center}
\caption{Types of coincidence backgrounds and number of 
expected coincidences in the SNO detector for the data set.
Upper limits and uncertainties on individual backgrounds
have been combined under the assumption that they are independent.}
\begin{tabular}{l r} \hline \hline
\multicolumn{2}{r}{ \nuebar \, background \hspace{2.cm}  expected} \\
Type of \nuebar &  coincidences \\ \hline 
\,\, Atmospheric  & 0.07 $\pm$ 0.01 \\
\,\, Reactor & 0.019 $\pm$ 0.002 \\ 
\,\, Diffuse supernovae & $\le$ 0.005 \\ 
\,\, geo-antineutrinos & 0.0 \\ \hline
Total \nuebar's \, background & 0.09 $\pm 0.01$  \\ \hline 

\multicolumn{2}{c}{ } \\
\multicolumn{2}{r}{ Non-\nuebar\, background \hspace{2.cm}  expected} \\
Process & coincidences \\ \hline
\,\, Atmospheric $\nu$ & 1.46 $^{+0.49}_{-0.45}$ \\
\,\,  $^{238}$U spontaneous fission in detector media & $<$ 0.79\\
\,\, Accidental coincidences & 0.13 $^{+0.06}_{-0.04}$ \\
\,\, $^{\mathrm{x}}$O(n,$\gamma$)$^{\mathrm{x+1}}$O , where x=17,18 & $<$ 0.05 \\
\,\, Instrumental contamination (95\% C.L.) & $<$ 0.027  \\ 
\,\, $^{13}$C($\alpha$,$n e^+ e^-$)$^{16}$O (90\% C.L.) & $<$ 1.7 $\times 10^{-3}$\\
\,\, Intrinsic: &  \\
\,\,\,\,\,\,\, $^{214}$Bi: $\beta\mbox{-}\gamma $ decay & 7.6$\times 10^{-5}$\\
\,\,\,\,\,\,\, $^{210}$Tl: $\beta\mbox{-}n$ decay & $\approx 10^{-8}$\\
\,\,\,\,\,\,\, $^{208}$Tl: $\beta\mbox{-}\gamma$ decay &  8.7$\times 10^{-4}$\\
\,\,\,\,\,\,\, $\gamma$ $\rightarrow$ Compton $e^-$ + photo-disintegration $n$ & $<$ 8 $\times 10^{-4}$ \\ \hline

Total non-\nuebar \, background & 1.59 $^{+0.93}_{-0.45}$ \\ \hline \hline

Total background & 1.68 $^{+0.93}_{-0.45}$  \\ \hline \hline
\end{tabular}
\label{tab_antinu_bkgd}
\end{center}
\end{table}

\section{Results}

The search for \nuebar \, candidates in the experimental data set
employs the same cuts on energy and fiducial volume as well as
the same coincidence extraction algorithms as
were used to derive the Monte Carlo-based coincidence detection efficiencies. 
One 3-fold and one 2-fold coincidence were found. 
Table \ref{tab_candid} summarizes the characteristics of the two \nuebar\, 
candidate coincidences and their constituent events. 
\begin{table}
\begin{center}
\caption{Two \nuebar\, candidate coincidences are found. Listed are kinetic 
recoil lepton energy and radial position for each constituent event as well as 
spatial separation and time separation relative to the first particle in each
coincidence.}
\begin{tabular}{ l r@{\ }l c c c c} \hline \hline
 \multicolumn{3}{l}{\nuebar\, candidate} & \
$T_{\mathrm{eff}}$ (MeV) \,\, &$r_{\mathrm{fit}} (\mathrm{cm})$ \,\, &
$\Delta r$ (cm)\,\, & $\Delta t$ (ms)\,\, \rule[0pt]{0pt}{8pt}\\ \hline

 I & 1st & particle & 8.58 & 283.2 & \phantom{88}0.0 & \phantom{8}0.0 \\
   & 2nd & particle & 5.39 & 472.4 & 206.7 & 16.7  \\
   & 3rd & particle & 5.15 & 349.2 & 178.3 & 20.3\\ \hline

 II & 1st & particle & 6.95 & 506.4 & \phantom{88}0.0 & \phantom{8}0.0 \\
    & 2nd & particle & 6.09 & 429.5 & \phantom{8}81.8 & 88.9 \\ \hline
\hline
\end{tabular}
\label{tab_candid}
\end{center}
\end{table}
On an event-by-event basis it is not possible to uniquely identify
individual constituent events as a positron or a neutron.
Therefore, the analysis re-groups ($e^+,n$)- and ($n,n$)-coincidences
into a single category of 2-fold coincidences. This category has an order
of magnitude higher sensitivity than 3-fold coincidences and 
is sensitive to \nuebar's\, with energies as low as 4~MeV.

\subsection{Differential Limits}
\begin{figure}[htb]
\includegraphics[width=3.4in]{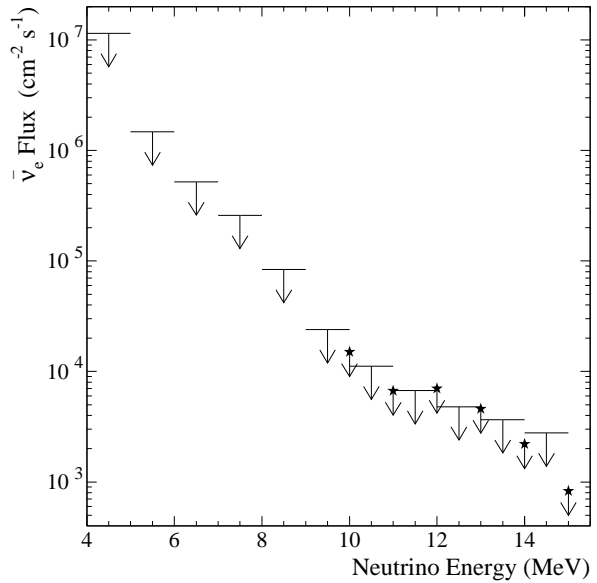}
\caption{Limits on the \nuebar \, flux from SNO (bars) and SK (stars). 
Bars represent 90\% C.L. flux limits for 1~MeV wide energy bins and are
 based on the assumption that the observed 2-fold coincidence originates from that particular energy bin. Stars indicate limits for mono-energetic \nuebar's .
}
\label{fig:abs_limits}
\end{figure}

This analysis sets model-independent differential limits on the \nuebar\,
flux in the neutrino energy range from 4 -- 14.8~MeV. Bin sizes in neutrino
energy were chosen to be 1~MeV. Based on the observed 2-fold coincidence, 
and under the conservative assumption of zero background, 
the Bayesian upper limit \cite{pdg96} on the number of 2-fold
coincidences amounts to 3.89 at the 90\% C.L.
In each neutrino energy bin, it is assumed that the candidate         
2-fold event was produced by an \nuebar\, of that energy.               
The upper limit on the number of candidate events is then         
corrected for detector acceptance and cross section to obtain         
a limit on the absolute $\bar{\nu}_e$ flux at that energy.         
As a result, the limit in each energy bin is model-independent         
and maximally conservative, but limits in different energy bins         
are strongly correlated.
Systematic uncertainties in the theoretical
cross sections, energy resolution differences between data and MC,
simulation failures, as well as systematics related to 
data reduction combine to about 2\% and have been taken into account.
Figure \ref{fig:abs_limits} displays \nuebar \, 
flux limits for the energy range from 4 -- 15~MeV 
at greater than 90\% C.L.
Super-Kamiokande's (SK) flux limits for mono-energetic \nuebar's\,
are shown for comparison \cite{gan03}. 
Super-Kamiokande's limits are based on data events, after subtraction of 
spallation background,  which fall in the $\pm 1 \sigma$
range of a Gaussian that describes the detector response to mono-energetic
\nuebar's. The SNO and SK limits are slightly different in nature since
SNO limits were calculated for a series of 1~MeV wide bins in neutrino 
energy.

\subsection{Integral Limit}

Under the assumption that the energy distribution of solar \nuebar's\, 
follows a $^8$B spectrum,  and that both observed candidates are of 
solar origin, an integral limit on the solar \nuebar\, flux is derived. 
The 2- and 3-fold coincidences are joined in order to maximize the sensitivity.
Using an extended Feldman-Cousins method \cite{fel98,con03} to include the 
background
uncertainty in the form of a two-sided Gaussian, the 90\% C.L. upper limit for 
2 candidate coincidences and 1.68 $^{+0.93}_{-0.45}$ background 
events has been calculated to be 3.8 coincidences. 
A MC calculation was used to convert a given \nuebar\, flux into a number of 
observed events. The 3.8 coincidences 
translate into a \nuebar\, flux limit of $\Phi <$ 3.4 $\times 10^4$cm$^{-2}$s$^{-1}$ in the energy range from 4 -- 14.8~MeV.
The systematic uncertainties have been treated similarly to the differential 
analysis.
The analysis energy window contains 83.4\% of the SSM-BP00 $^8$B \nue\, 
flux of 5.05$ ^{+1.01}_{-0.81} \times 10^6$cm$^{-2}$s$^{-1}$ \cite{bp00}.
The above limit implies a 90\% C.L. upper bound on the 
conversion probability of solar $^8$B $\nu_e$'s into \nuebar's\, of 0.81\%,
if \nuebar's \, are assumed to follow a $^8$B spectrum.
This assumption is equivalent to an 
energy-independent $\nu_e \rightarrow \overline{\nu}_e$  conversion 
mechanism.\\
\indent If the analysis is restricted to the \nuebar\, energy range from 4 -- 8~MeV,
only the observed 2-fold coincidence represents a \nuebar\, candidate since
the 3-fold coincidence could only have originated from a \nuebar\, with an 
energy in excess of 12.6~MeV. Within the 4 -- 8~MeV energy window the 
background is conservatively assumed to be zero coincidences. Using a Bayesian
prescription \cite{pdg96}, the 90\% C.L. upper limit on one candidate 
and zero assumed background corresponds to 3.89 events.
Assuming a fission spectral shape \cite{vog81,sch89} from possible naturally 
occurring elements, this defines a limit of
$\Phi <$ 2.0$\times 10^6$cm$^{-2}$s$^{-1}$ in the energy 
range from 4 -- 8~MeV which covers 9\% of the average reactor \nuebar\, flux.

\subsection{Comparison with other Experiments}
Previously, other experiments have set very stringent limits on the
\nuebar \, flux from the Sun.
Under the assumption of an unoscillated $^8$B spectral shape for 
solar \nuebar's \,
KamLAND \cite{egu04} limits the solar flux of \nuebar's\, to less than
3.7 $\times$ 10$^2$ cm$^{-2}$s$^{-1}$ (90\% C.L.).
This measurement is based on the neutrino energy range from 8.3 -- 14.8~MeV
and corresponds to an upper limit on the $\nu_e \rightarrow$ \nuebar\, 
conversion probability of 2.8 $\times$ 10$^{-4}$ (90\% C.L.).\\
\indent SK's integral flux limit is based on the energy range from 8 -- 20~MeV and, 
under the assumption of a $^8$B spectrum for solar \nuebar's and a total 
\nuebar\, flux of 5.05 $\times$ 10$^6$ cm$^{-2}$s$^{-1}$, 
places a 90\% C.L. limit on the conversion probability of less than 
0.8\% \cite{gan03}. This corresponds to an absolute flux limit of 
1.4 $\times$ 10$^4$ cm$^{-2}$s$^{-1}$ in the energy region from 8 -- 20~MeV
which contains 34.4\% of the total $^8$B flux.
The present SNO analysis investigates the energy range from 4 -- 14.8~MeV and 
uses a completely independent direct counting method with very low background.
Therefore it can set a comparable limit on the \nuebar\, flux 
despite the fact that SK's exposure is 800 times larger.
Due to the time dependence of the solar magnetic field the \nuebar\, flux
originating from conversion of solar $^8$B neutrinos could vary as function
of time. Table \ref{tab_sol_lim} specifies the existing limits on the 
\nuebar\, flux and indicates the time frames during which the relevant data
were recorded.
\squeezetable
\begin{table}
\begin{center}
\caption{Integral \nuebar\, flux limits \nuebar/SSM \nue\, at the 90\% C.L. relative to the SSM-BP00 $^8$B flux \cite{bp00}, periods of time during which the relevant data were taken, and energy ranges over which the various experiments
have sensitivity to solar \nuebar\, are presented.} 
\begin{tabular}{ l l r r } \hline \hline
Experiment & Time Period & Energy (MeV) & Limit(\%) \\ \hline
KamLAND \cite{egu04} & 3/4/2002 -- 12/1/2002 & 8.3 -- 14.8 & 0.028 \\
SNO & 11/2/1999 -- 5/28/2001 & 4.0 -- 14.8 & 0.81 \\
SK \cite{gan03} & 5/31/1996 -- 7/15/2001 & 8.0 -- 20.0 & 0.8 \\
LSD \cite{agl96} & before 4/1996 & 8.8 -- 18.8 & 1.95  \\
Kamiokande \cite{ino93} & 6/1988 -- 4/1990 & 12.0 -- 13.0 & 5.07 \\
\hline
\end{tabular}
\label{tab_sol_lim}
\end{center}
\end{table}\\
\indent The present analysis provides a 90\% C.L. upper limit for \nuebar\, energies
from 4 -- 8~MeV of 2.0$\times$10$^{6}$~cm$^{-2}$s$^{-1}$ assuming a
neutron induced fission spectral shape. 
Because the contribution from man-made reactors in the
region of the SNO detector is calculated to be very low, (see Table \ref{tab_antinu_bkgd}),
this can be considered as an upper limit on the flux from naturally
occurring neutron induced fission sources. 
However, we note that if all 54 events
observed by KamLAND \cite{egu03}  for \nuebar\, energies from 3.4 -- 8~MeV were
considered to originate from the neutron induced fission of 
naturally-occurring elements rather than from nearby man-made reactors, 
an upper limit on the \nuebar\, flux in the 3.4 -- 8~MeV energy region 
of 1.55 $\times$ 10$^5$ cm$^{-2}$s$^{-1}$ (90\% C.L.) 
can be derived for the KamLAND location. During the completion of this paper,
KamLAND \cite{egu03} reported a new result of 258 events in a 515-day live 
time and a 33\% larger fiducial volume. 
No significant change in the \nuebar\, flux limit is expected as a result.\\
\indent The neutron detection efficiency in pure D$_2$O was 
14.4\% \cite{snonc}.  
For the phase of the SNO experiment in which NaCl was
added to the D$_2$O, improving the neutron detection efficiency to 39.9\% 
\cite{sno03}, an increase in the effective \nuebar\, sensitivity
by a factor of about 3 is anticipated.\\

\subsection{Neutrino Decay}

Although flavor transformation of solar neutrinos is
believed to be dominated by matter-enhanced neutrino
oscillations with mixing parameters in the LMA
region, SFP mechanisms or neutrino decay
may also contribute. 
Since neutrino decay is expected to be energy-dependent,
SNO's low \nuebar\, energy threshold of 4~MeV is a valuable
feature to test for non-radiative neutrino decay of the form
$\nu_2 \rightarrow \bar\nu_1 + X$.
Here $\nu_2$ and $\nu_1$ refer to the heavier and lighter 
neutrino mass eigenstates and $X$ is a scalar particle
(e.g. a Majoron) \cite{nudec}.
For quasi-degenerate neutrino masses, a lower limit on 
the lifetime $\tau_2$ of the heavier neutrino is found to be
$\tau_2/m_2 >$0.004 s/eV. For hierarchical neutrino masses,
the limit amounts to $\tau_2/m_2 >$  4.4 $\times 10^{-5}$ s/eV,
equivalent to $\tau_2 >$ 0.44~$\mu$s if $m_2 \approx$ 0.01 eV \footnote{We 
thank S. Pakvasa and J. Busenitz for valuable discussions on neutrino decay.}.
Previously, KamLAND \cite{egu04} has presented lower limits on non-radiative 
neutrino decay based on the \nuebar\, energy range from 8.3 -- 14.8~MeV,
and found 
$\tau_2/m_2 >$0.067 s/eV for quasi-degenerate and $\tau_2 >$ 11~$\mu$s 
for hierarchical neutrino masses.

\section{Conclusion}

In summary, our analysis represents a novel detection technique 
to search for \nuebar's, with very low backgrounds. 
Based on the one 2-fold and one 3-fold observed
coincidence, integral limits on the \nuebar\, flux in the energy range
below 8~MeV and in the range from 4 -- 14.8~MeV have been set under the 
assumption of a fission and a $^8$B spectrum, respectively. 
Spectrally independent differential limits have been placed as well.
The derived limit on the flux of solar \nuebar's \, was used to constrain
the neutrino lifetime.
Within SNO's sensitivity we independently confirm the SK and KamLAND results
on \nuebar\, fluxes.

\section*{ACKNOWLEDGMENTS}
This research was supported by:  Canada: Natural Sciences and Engineering Research Council, Industry Canada, National Research Council,
Northern Ontario Heritage Fund Corporation, Inco, Atomic Energy of Canada, Ltd., Ontario Power
Generation; US: Dept. of Energy; UK: Particle Physics and Astronomy Research Council. We thank the SNO technical
staff for their strong contributions.

\end{document}